\documentclass[aps,twocolumn,floats]{revtex4-2}
\usepackage{amsmath}
\usepackage{graphicx}
\usepackage{dcolumn}
\usepackage{bm}
\usepackage{amssymb}
\usepackage{latexsym}
\usepackage{color}
\usepackage{subfigure}
\usepackage[normalem]{ulem}
\usepackage{cancel}
\usepackage{comment}

\begin{document}

\title{Fermionic atoms in a spin-dependent optical lattice potential: topological insulators with broken time-reversal symmetry}

\author{Igor Kuzmenko$^{1}$,  Miros{\l}aw Brewczyk$^2$, Grzegorz {\L}ach$^3$, Marek Trippenbach$^3$, Y. B. Band$^1$}

\address{$^1$ Department of Chemistry, Department of Physics,
Department of Electro-Optics, and the Ilse Katz Center for Nano-Science,
Ben-Gurion University, Beer-Sheva 84105, Israel 
\\
$^2$ Wydzia{\l} Fizyki, Uniwersytet w Bia{\l}ymstoku,  ul. K. Cio{\l}kowskiego 1L, 15-245 Bia{\l}ystok, Poland
\\
$^3$ Faculty of Physics, University of Warsaw, ul. Pasteura 5, 02-093 Warsaw, Poland
}

\begin{abstract}
We propose a  novel approach to study the topological properties of matter.  In this approach, fermionic atoms are placed in an external magnetic field and in a two-dimensional spin-dependent optical lattice (SDOL) created by intersecting laser beams with a superposition of polarizations.  To demonstrate the utility of the SDOL-based technique we compute the topological invariants (Chern numbers) for the SDOL bands as a function of an external magnetic field, and show the existence of a rich topology of the energy bands for this system which does not have parity-time-reversal symmetry. We explicitly consider $^{6}$Li $F=1/2$ atoms.  Using a projection matrix method we observe topological phase transitions between an ordinary insulator, an abelian topological insulator, and a non-abelian topological insulator as the external magnetic field strength is varied. Upon introducing edges for the SDOL we find topological edge states (that are correlated with the band Chern numbers) that simultaneously exhibit non-trivial density and spin currents with both a rotational flow contribution and flow along the edge of the SDOL.
\end{abstract}

\maketitle

\section{Introduction}  \label{sec:Intro}

A topological insulator is a material whose interior is unable to conduct an electric current while its edges support such a flow \cite{Hasan_Kane_10, Asboth, Qi_Zhang_11, Cooper_19, Goldman_14}.  A topological insulator (TI) differs from an ordinary insulator in that it is not possible to continuously deform spin-orbit-induced topological insulators into an ordinary one without closing the bulk gap, i.e., without undergoing a topological phase (TP) transition. The presence of time-reversal symmetry is crucial for that \cite{Chiu_16}. TIs have been theoretically studied \cite{Asboth, Hasan_Kane_10, Bernevig_06} and experimentally realized in a variety of systems, including HgTe/CdTe semiconductor quantum wells \cite{Konig_07}, BiSb alloys \cite{Hsieh_08}, and Bi$_2$Se$_3$ crystals \cite{Hsieh_09,Xia_09}. 

Significant progress has been made in the realization of band structures with non-trivial topology in ultracold atomic gas experiments \cite{Cooper_19, Goldman_14}. The use of cold atoms in the study of TPs of matter offers the advantage of well-controlled experimental parameters.  Furthermore, recent advances have also been made in the generation of fictitious magnetic fields and spin-orbit coupling for ultracold neutral atoms in optical potentials (see Refs.~\cite{Zhai_15, Dalibard_16, Aidelsburger_18, Wu_16, Huang_16}).

In this paper we propose another promising method for the study of topological properties of matter that does not possess parity-time-reversal symmetry. In contrast to existing ultracold-atom methods, which are based on periodically shaken lattices \cite{Struck_13, Jotzu_14, Flaschner_16} or Raman coupling of internal states \cite{Stuhl_15, Mancini_15}, our approach involves placing  atoms in a two-dimensional spin-dependent optical lattice potential and an external magnetic field. A simplified study of fermionic and bosonic atoms in a spin-dependent optical lattice (SDOL) in the limit of singly occupied sites was recently carried out in Refs.~\cite{Kuzmenko_19, Szulim-SDOLP-2022}. Here we use a Bloch band model that goes beyond the two-band tight-binding approximation \cite{Mao-tight-binding-topological-PRB-2011} to describe the SDOL potential, which allows us to determine the topology of the high-energy bands, which exhibit multiple tangled bulk band gaps \cite{Jiang_21}. We find a sequence of TP transitions upon increasing the external magnetic field strength, involving abelian and non-abelian TPs. The rich topology of the high-energy bands is also reflected in the characteristics of the edge states.

Below we consider a cold fermionic gas of $^{6}$Li $F= 1/2$ atoms in a SDOL and an external magnetic field.  We show that the properties of the system are significantly enriched relative to the system without external magnetic field, and that a radical change of the topological properties can be observed as the strength of the external magnetic field is varied. We calculate the band-structure and the Chern numbers (the topological invariants that classify bands in topological materials \cite{Hasan_Kane_10, Asboth,  Qi_Zhang_11, Goldman_14, Cooper_19}) of the bands. Then, we apply blue-detuned lasers which introduce edges to the SDOL, and we investigate their topological character.

The paper is organized as follow:
Section~\ref{sec:SDOLP} introduces the Hamiltonian of atoms trapped in
a two-dimensional spin-dependent optical lattice potential (SDOLP) in the presence of an external magnetic field perpendicular to the SDOLP plane.
Section~\ref{sec:Chern_nos} calculates the Chern numbers, denoted by $C_n$ of Bloch bands,
which are parameterized by an integer band number $n$.
In Sec.~\ref{sec:Edge_sts}, energies and wave functions of edge states are calculated.
The numerical calculation of the energies and wave functions of trapped atoms is performed for a variety of external magnetic field strengths.
Section~ \ref{sec:Non-abelian_phase} presents a generalization of the concept of an eigenvector frame rotation in a non-abelian topological phase employing the projection matrices.
The transition from abelian to non-abelian is indicated by a discontinuity in the Frobenius norm of the projection matrix.
In Section~\ref{sec:density_curr-density_spin-curr-density}, we analyze the 
atom probability density, spin density, current density, and spin-current density in order to characterize the edge states.
A summary of the results obtained is provided in Section~\ref{sec:summary}.
The appendices contain a number of technical details. 
Appendix~\ref{append:SDOLP-symmetries} addresses
the symmetries of the SDOLP Hamiltonian in both
the absence and presence of an external magnetic field.
The TP transitions observed in the presence of an external magnetic field
are described in Appendix~\ref{append:TPTs}.
In Appendix~\ref{append:CN} a detailed calculation of the Chern numbers is presented, while
edge state degeneracy is discussed in  Appendix~\ref{sec:edge-states}.

\section{Spin-Dependent Optical Lattice Potential}  \label{sec:SDOLP}

The spin-dependent optical lattice potential is generated by two pairs of counter-propagating linearly polarized laser beams, which are tightly bound in the $z$-direction and form two-dimensional lattice.  The polarization of the beams is varied such that the complex slowly varying envelope ${\bf E}({\bf r})$ of the electric field is given by
\begin{eqnarray}
  {\boldsymbol {\mathcal E}}({\bf r},t) &=&
  \frac{1}{2} ({\bf E}({\bf r}) e^{-i\omega_l t} + {\rm{c.c.}}) ,
  \label{eq:ElectricField-real}
  \\
  {\bf E}({\bf r}) &=&
  \frac{E_0 }{\sqrt{2}}
  \sum_{n=1}^4
  \left( {\hat {\bf z}} + {\hat {\bf k}}_n \times {\hat {\bf z}} \right) \,
  e^{i {\bf k}_n \cdot {\bf r}} ,
  \label{eq:ElectricField-complex}
\end{eqnarray}
where ${\bf r} = (x,y)$, ${\hat {\bf z}}$ is the polarization unit vector along the $z$-axis, the wavevectors are
\begin{equation}
  {\bf k}_n =
  k_l \,
  \bigg(
    \cos \Big[ \frac{(2n+1) \pi}{4} \Big] , \,
    \sin \Big[ \frac{(2n+1) \pi}{4} \Big] , \, 0
  \bigg) \,,
\end{equation}
$k_l = 2\pi/\lambda_l$ is the laser wavenumber, $\lambda_l$ is the laser wavelength, and ${\hat {\bf k}}_n$ is the unit vector in the direction of ${\bf k}_n$.

The total atomic Hamiltonian in the SDOLP can be written as
\begin{equation}   \label{eq:Hamiltonian}
  H = -\frac{\hbar^2}{2 M} \, \nabla^2 +  H_{\mathrm{Stark}}({\bf r}) + H_Z .
\end{equation}
The first term on the right hand side of Eq.~(\ref{eq:Hamiltonian}) is the kinetic energy operator of an atom, where $M$ is the atomic mass.  The last term on the right hand side of Eq.~(\ref{eq:Hamiltonian}), $H_Z = -g_F \mu_B \, B_{\mathrm{ext}} F_z$, is the Zeeman interaction Hamiltonian of the atom with an external magnetic field ${\bf B}_{\mathrm{ext}} = B_{\mathrm{ext}} {\hat {\bf z}}$.  The middle term, $H_{\mathrm{Stark}}({\bf r})$, is the optical lattice Stark interaction Hamiltonian.  Some symmetries of the SDOLP Hamiltonian in both the absence and presence of an external magnetic field are discussed in Appendix~\ref{append:SDOLP-symmetries}.

For $^{6}$Li, the total electronic angular momentum is $J = 1/2$, hence, the effective interaction of an atom with the electromagnetic field can be described using a scalar potential $V$ and vector potential containing what Cohen-Tannoudji and Dupont-Roc termed the fictitious magnetic field ${\bf B}_\mathrm{fic}$ \cite{Kuzmenko_19,SOI-EuroPhysJ-13, Cohen-Tannoudji_72, Dudarev-PRL-2004}, and the tensor potential vanish. The Stark Hamiltonian in the presence of a fictitious magnetic field is given by
\begin{equation}   \label{eq:H-V+B}
  H_{\mathrm{Stark}} (\mathbf{r}) =
  V (\mathbf{r}) -
  g_F \mu_B \, \mathbf{B}_{\mathrm{fict}} (\mathbf{r}) \cdot \mathbf{F} \,.
\end{equation}
Here ${\bf F}$ is the atomic hyperfine angular momentum, $\mu_B$ is the Bohr magneton, and for a ground state alkali atom with $L = 0$, $J = S = 1/2$, $I = 1$ nuclear spin, $g_F = g_S \, \tfrac{F (F+1) - I (I+1) + S (S+1)}{2 F (F+1)}$, and $g_S$ is the electron spin $g$-factor.  The SDOLP is defined as follows $V({\bf r}) = -\tfrac{\alpha_s(\omega_l)}{4} \, {\bf E}^*({\bf r}) \cdot {\bf E}({\bf r})$ and ${\bf B}_\mathrm{fic}({\bf r}) = \tfrac{i}{2 F} \, \tfrac{ \alpha_v(\omega_l)}{4 g_F \mu_B} \, {\bf E}^*({\bf r}) \times {\bf E}({\bf r})$, where $\alpha_s$ and $\alpha_v$ are scalar and vector polarizabilities which depend upon the detunings of the laser frequency $\omega_l$ from the D$_1$ and D$_2$ resonance lines of Li. Explicitly,
\begin{eqnarray}
  V({\bf r} ) &=&
  -\frac{V_0}{2} \, \big[ 2 + \cos \big( q_0 x \big) + \cos \big( q_0 y \big) \big] ,
  \\
  {\bf B}_\mathrm{fic}({\bf r}) &=&
  - B_0 {\hat {\bf x}} \, \sin \big( q_0 x \big) \, \cos^2 \big( \frac{q_0 y}{2} \big)
  \nonumber \\ &&
  - B_0 {\hat {\bf y}} \, \sin \big( q_0 y \big) \, \cos^2 \big( \frac{q_0 x}{2} \big) .
\end{eqnarray}
Here $V_0 = \alpha_s(\omega_l) E_0^2$, $B_0 = \tfrac{\sqrt{2}}{2 F g_F \mu_B} \alpha_v(\omega_l)  E_0^2$ and $q_0 = \sqrt{2} \, k_l$. 
Note that the divergence of the fictitious magnetic field does not vanish; ${\bf B}_\mathrm{fic}({\bf r})$ corresponds to a radially distributed magnetic monopole density.  The Hamiltonian (\ref{eq:Hamiltonian}) has square lattice symmetry with lattice vectors ${\bf a}_1 = (a_0, 0)$ and ${\bf a}_2 = (0, a_0)$, where the lattice period is $a_0 = 2\pi /q_0$, hence the reciprocal lattice has square symmetry with reciprocal lattice vectors ${\bf q}_1 = (q_0,0)$ and ${\bf q}_2 = (0,q_0)$.

A ${}^{6}$Li atom in the ground state has $F = 1/2$, and the scalar potential $V(\mathbf{r})$ and the fictitious magnetic field $\mathbf{B}_{\mathrm{fic}} (\mathbf{r})$ in the vector potential are asigned as previously described, with polarizabilities $\alpha_{s} (\omega_l) =\alpha_{n J F}^{s} (\omega_l)$ and $\alpha_{v} (\omega_l) =\alpha_{n J F}^{v} (\omega_l)$ defined by
\cite{SOI-EuroPhysJ-13}
\begin{eqnarray}
  \alpha_{n J F}^{s} (\omega_l) &=&
  \frac{1}{\sqrt{3 (2 J + 1)}} \,
  \alpha_{n J}^{(0)} (\omega_l) ,
  \label{eq:alpha_s}
  \\
  \alpha_{n J F}^{v} (\omega_l) &=&
  \big( -1 \big)^{J + I + F}
  \sqrt{\frac{2 F (2 F + 1)}{F + 1}}
  \nonumber \\ && \times
  \left\{
    \begin{array}{ccc}
      F & 1 & F
      \\
      J & I & J
    \end{array}
  \right\} \,
  \alpha_{n J}^{(1)} (\omega_l) ,
  \label{eq:alpha_v}
\end{eqnarray}
where $\left\{ \begin{array}{ccc} F & 1 & F \\ J & I & J \end{array} \right\}$
are the Wigner 6-$j$ symbols.
The reduced scalar and vector polarizabilities are
\begin{eqnarray}
  \alpha_{n J}^{(K)} (\omega_l) &=&
  \frac{1}{\hbar} \,
  \sqrt{2 K + 1}
  \sum_{n' J'}
  \big( -1 \big)^{K + J + J' + 1}
  \nonumber \\ && \times
  \left\{
    \begin{array}{ccc}
      1 & K & 1
      \\
      J & J' & J
    \end{array}
  \right\} \,
  \big|
    \langle n' J' \| d \| n J \rangle
  \big|^{2}
  \nonumber \\ && \times
  \mathrm{Re}
  \bigg[
    \frac{1}{\omega_{n' J' n J} - \omega_l - \frac{i}{2} \gamma_{n' J' n J}}
    \nonumber \\ && +
    \frac{(-1)^{K}}{\omega_{n' J' n J} + \omega_l + \frac{i}{2} \gamma_{n' J' n J}}
  \bigg] ,
\end{eqnarray}
where $\langle n' J' \| d \| n J \rangle$ is reduced matrix element of
the electric dipole transition $n J \to n' J'$,
$\omega_{n' J' n J}$ is the transition frequency, and
$\gamma_{n' J' n J}$ is the transition line-width.

The ground electronic state of the lithium atom is the $2 \, {}^{2}S_{1/2}$ state, and the excited electronic states are the $2 \, {}^{2}P_{1/2}$ state and the $2 \, {}^{2}P_{3/2}$ state.  The $D_1$ transition, $2 \, {}^{2}S_{1/2} \to 2 \, {}^{2}P_{1/2}$, has the transition frequency $\omega_{D_1} \equiv \omega_{2 \frac{1}{2} 2 \frac{1}{2}} = 2 \pi \times 446.789 634$~THz \cite{res_freq}, the linewidth is $\gamma_{D_1} \equiv \gamma_{2 \frac{1}{2} 2 \frac{1}{2}} = 2 \pi \times 5.8724$~MHz \cite{res_width} and the reduced matrix element $d_{D_1} \equiv \langle 2 \frac{1}{2} \| d \| 2 \frac{1}{2} \rangle = - 8.433 \times 10^{-18}~\text{esu} \times \text{cm}$.  The $D_2$ transition, $2 \, {}^{2}S_{1/2} \to 2 \, {}^{2}P_{3/2}$, has the transition frequency $\omega_{D_2} \equiv \omega_{2 \frac{3}{2} 2 \frac{1}{2}} = 2 \pi \times 446.799 677$~THz  \cite{res_freq}, linewidth $\gamma_{D_2} \equiv \gamma_{2 \frac{3}{2} 2 \frac{1}{2}} = 2 \pi \times 5.8724$~MHz  \cite{res_width} and reduced matrix element $d_{D_2} \equiv \langle 2 \frac{3}{2} \| d \| 2 \frac{1}{2} \rangle = 11.925 \times 10^{-18}~\text{esu} \times \text{cm}$.

\begin{figure}[thb] 
\includegraphics[width=5.5cm]{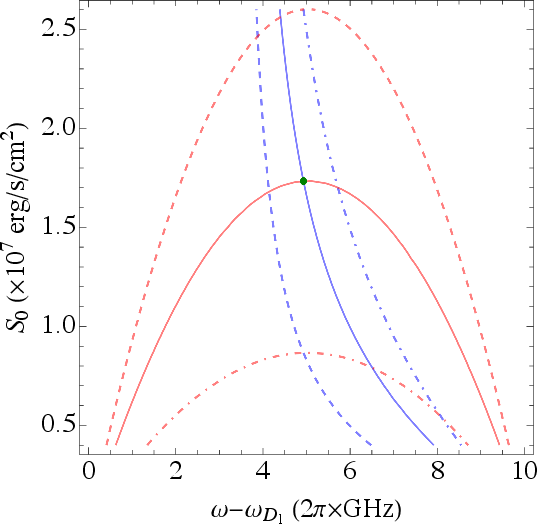}
\caption{The electromagnetic energy flux per unit area $S_0$ versus detuning from the $D_1$ line. The curves of equal $V_0$ (blue):
  $V_0 = 2.5 \, \mathcal{E}_{0}$ (dashed),
  $V_0 = 5 \, \mathcal{E}_{0}$ (solid) and
  $V_0 = 7.5 \, \mathcal{E}_{0}$ (dot-dashed),
  where $\mathcal{E}_{0}$ is the recoil energy.
  The curves of equal $B_0$ (red):
  $B_0 = 5 \, \mathcal{B}_{0}$ (dot-dashed),
  $B_0 = 10 \, \mathcal{B}_{0}$ (solid), and
  $B_0 = 15 \, \mathcal{B}_{0}$ (dashed),
  where $\mathcal{B}_{0} = \mathcal{E}_{0} / (g_F \mu_B)$.
  The green dot is a crossing point of the blue and red solid curves.}
\label{Fig:V0-B0-vs-Felta-S0}
\end{figure}

Note that $V_0$ and $B_0$, which are proportional to the scalar polarizability $\alpha_s(\omega_l)$ and the vector polarizability $\alpha_v (\omega_l)$ respectively, depend on the laser frequency $\omega_l$ and the electromagnetic energy flux per unit area $S_0 = \frac{c}{4 \pi} E_{0}^{2}$  (in Gaussian units).  Figure~\ref{Fig:V0-B0-vs-Felta-S0} shows the curves of the constant $V_0$ and the curves of the constant $B_0$, where the $x$-axis is the detuning of the laser frequency $\omega$ from the resonance frequency $\omega_{D_1}$ of the $D_1$ transition, and the $y$-axis is the electromanetic energy flux per unit area $S_0$. The green dot is the point $\omega = \omega_{D_2} - 2 \pi \times 5.11678$~GHz and $S_0 = 1.73328 \times 10^7$ ~ erg s$^{-1}$ cm$^{-2}$, where $V_0 = 5 \, \mathcal{E}_{0}$ and $B_0 = 10\, \mathcal{B}_{0}$, where $\mathcal{E}_{0} = \frac{\hbar^2 \omega_{l}^{2}}{2 M c^2}$ is the recoil energy, $\mathcal{B}_{0} = \mathcal{E}_{0} / (g_F \mu_B)$. Thus, the laser frequency $\omega_l$ which is used is red-detuned from the Li  D$_2$ line by $\Delta = - 2\pi \times 5.117$~GHz.

\begin{figure}
\centering
  \includegraphics[width=\linewidth,angle=0] {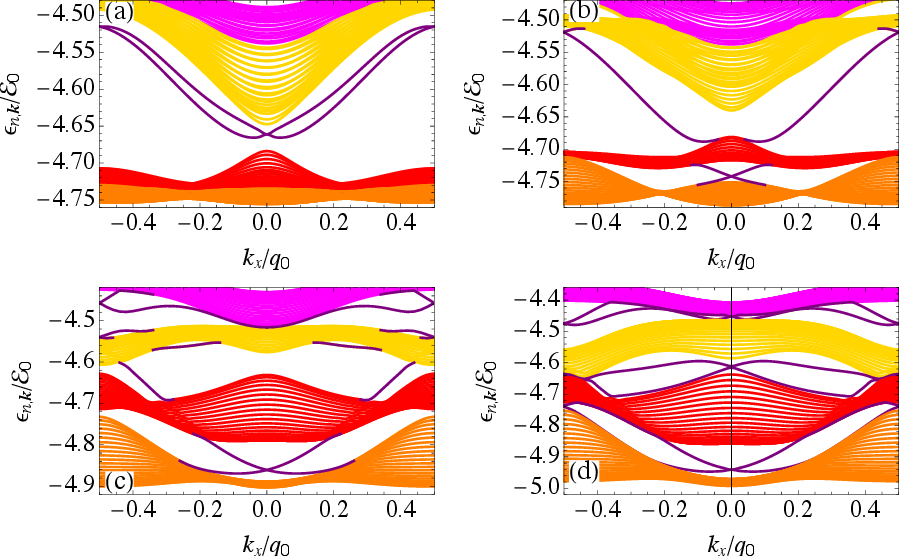}
\caption{\footnotesize  
  Band energies $\epsilon_{n, k_x}$ versus $k_x$ numbered by
  $n = 5$ (orange), $n = 6$ (red), $n = 7$ (gold), $n = 8$ (magenta),
  for $V_0 = 5 \, \mathcal{E}_{0}$, $B_0 = 10 \, \mathcal{B}_{0}$, and
  four values of the external magnetic field $B_{\mathrm{ext}}$:
  (a) $B_{\mathrm{ext}} = 0.05 \, \mathcal{B}_{0}$,
  (b) $B_{\mathrm{ext}} = 0.2 \, \mathcal{B}_{0}$,
  (c) $B_{\mathrm{ext}} = 0.7 \, \mathcal{B}_{0}$, and
  (d) $B_{\mathrm{ext}} = 1.0 \, \mathcal{B}_{0}$.  Also shown
  are the edge-state energies (purple) that result when the blue-detuned
  potential $V_{\mathrm{BD}}(y) (y) = V_{\mathrm{BD},0}  \Theta (|y| - L_y/2)$ is introduced where $V_{\mathrm{BD},0}$ is
  taken to be infinite and $L_y = 25 a_0$. To experimentally probe high-lying bands
  and edge states, one would fill the Fermi sea to a desired energy.}
\label{Fig:energy-NDEigensystem-vs-Bext}
\end{figure}

The quantum states of the atoms in the SDOLP are parametrized by wavevector ${\bf k} = (k_x, k_y)$ belonging to the first Brillouin zone (BZ) of the SDOLP: $| k_x, k_y | < q_0/2$, and the energy band number $n$, where $n$ is a positive integer.  The corresponding energies $\epsilon_{n, {\bf k}}$ and wave functions $\psi_{n, {\bf k}}({\bf r})$ are determined from the Sch\"odinger equation $H \psi_{n, {\bf k}}({\bf r}) = \epsilon_{n, {\bf k}} \psi_{n, {\bf k}}({\bf r})$.  The Sch\"odinger equation was solved numerically using the {\it Mathematica} command NDEigensystem to find radial wave functions $\psi_{n, {\bf k}}({\bf r})$ and eigenenergies $\epsilon_{n, {\bf k}}$.  Figure~\ref{Fig:energy-NDEigensystem-vs-Bext} shows the bands calculated for specific values of $V_0$, $B_0$, and $B_{\mathrm{ext}}$ as specified in the figure caption (energy and magnetic field are in units of $\mathcal{E}_{0}$ and $\mathcal{B}_{0}$, respectively).  The figure also shows edge states in a finite width strip produced by blue-detuned lasers, see below.

The scalar potential $V(\mathbf{r})$ has minima at $\mathbf{r} = \mathbf{a} \equiv n_1 \mathbf{a}_{1} + n_2 \mathbf{a}_{2}$ where $n_1$ and $n_2$ are integers.  ${\bf B}_\mathrm{fic}({\bf r})$ vanishes at $\mathbf{r} = \mathbf{a}$, and at the edges of the Wigner-Seitz cells, $\mathbf{r} = \mathbf{a} + \tfrac{1}{2} \mathbf{a}_{1}$ and $\mathbf{r} = \mathbf{a} + \tfrac{1}{2}  \mathbf{a}_{2}$.  Hence the minimum of $V(\mathbf{r})$ at 
$\mathbf{a}$ [where $V(\mathbf{a}) = -2 V_0$] and the nearest-neighbor minima at $\mathbf{a} + {\bf a}_j$ ($j = 1, 2$) are separated by a barrier of height $V_0$.
Hence, the bands with $\epsilon_{n, \mathbf{k}} < -V_0$ can be well described using a tight-binding model. However, for the bands which are above the barriers, $\epsilon_{n, {\bf k}} > -V_0$, tight-binding is a poor approximation.  All bands shown in Fig.~\ref{Fig:energy-NDEigensystem-vs-Bext} are above $-V_0$, hence tight-binding cannot be used for them. Note that for  $^{6}$Li, singlet $s$-wave scattering length is small \cite{Abraham_96} and a single particle picture can be used.

\section{Chern numbers}   \label{sec:Chern_nos}

Chern numbers $C_n$ for the Li atoms in the SDOLP are defined on the bands in wavevector space \cite{Hasan_Kane_10, Cooper_19}. For a 2D periodic system, $C_n$ for the $n$th band is determined by integrating the Berry curvature, $\mathcal{F}_{xy}(n,{\bf k}) = \partial_{k_x} A_y(n,{\bf k}) - \partial_{k_y} A_x(n,{\bf k})$, over the first BZ,
\begin{equation}\label{chern-number}
C_n=\frac{1}{2\pi} \int_{\rm BZ} d {\bf k} \left( \frac{\partial A_y(n,{\bf k})}{\partial k_x} - \frac{\partial A_x(n,{\bf k})}{\partial k_y}\right) ,
\end{equation}
where $A_\mu(n,{\bf k}) = -i \langle n, {\bf k} | \frac{\partial}{\partial k_\mu} |n, {\bf k} \rangle$ is the Berry connection. 
In our analysis we used an efficient numerical method proposed in Ref.~\cite{Fukui_05}. The details are presented in Appendix~\ref{append:CN}.

Increasing the external magnetic field strength we find a sequence of TP transitions. Phase transitions occur when bandgaps between Bloch bands close and open. The energy gap can vanish at several points in the BZ: the $\Gamma$ point, the vertices of the BZ, and the edge centers of the BZ.  Some of the gapless spectra have Dirac cones and some have quadratic wavevector dependence. The summary of all TP transitions found for external magnetic fields $B_{\mathrm{ext}} < 2\, \mathcal{B}_{0}$ is given in Appendix~\ref{append:TPTs}.  The lowest critical points at which the TP transitions occur are $\{ B_{c,1}, B_{c,2}, B_{c,3}, B_{c,4}, B_{c,5} \} = \{ 0.11, 0.66, 0.89, 1.25, 1.35 \}\mathcal{B}_{0}$. All the Chern numbers vanish below $B_{c,1}$.  At the lowest TP transition the gap between the $6$th and $7$th bands closes at the center of the BZ and forms a Dirac cone, see Fig.~\ref{DiracCone} in Appendix~\ref{append:TPTs}.  For $B_{c,1} < B_{\mathrm{ext}} <  B_{c,2}$, $C_6 = -1$ and  $C_7 = 1$. Indeed, in this case the bulk-boundary correspondence leads to the appearance of two topologically protected chiral edge states in the bandgap, see discussion below. Above $B_{c,2}$ the second TP transition appears and the gap between $7$th and $8$th energy bands closes, again at ${\bf k} = {\bf 0}$, and $C_7$ changes from $+1$ to $-1$, and $C_8$ changes from $0$ to $+2$.  Note that the sum of all the Chern numbers in the eight lowest energy bands is always equal to zero.

Since the Chern numbers $C_n$ are non-zero, this system is a TI.   For abelian TIs, the Chern numbers are such that $C_n = - C_{n+1}$, and the number of chiral edge states on each edge between bands $n$ and $n+1$ is $|C_n|$ \cite{Hasan_Kane_10,Asboth}. For non-abelian TIs the bulk-edge correspondence is more subtle than the abelian case \cite{Jiang_22}; multiple tangled bulk bandgaps are present and the system supports non-trivial edge states.

\section{Edge states}  \label{sec:Edge_sts}

In order to study edge states for the atoms in the SDOLP we introduce a blue-detuned potential of the form $V_{\mathrm{BD}}(y) = V_{\mathrm{BD},0} \Theta (|y| - L_y/2)$ where $\Theta(\bullet)$ is the Heaviside Theta function.  This potential mimics the effects of a blue-detuned laser that repels atoms from the region $|y| > L_y/2$. For convenience we take $V_{\mathrm{BD},0}$ very large so we can apply Dirichlet boundary conditions at the edge. To compute the edge states we again use the {\it Mathematica} command NDEigensystem with Dirichlet boundary conditions at $y = \pm L_y/2$ and periodic boundary conditions in $x$.  

Figure~\ref{Fig:energy-NDEigensystem-vs-Bext} shows the edge states of the finite-width strip, $|y| \le L_y/2$, as purple curves that lie within the bandgaps (and within the bands) of the fully periodic system. Not all the edge states shown in purple in Fig.~\ref{Fig:energy-NDEigensystem-vs-Bext} are topological edge states (TESs).  Clearly those that connect adjacent bands and lie in the gap between them are TESs. The edge states in Fig.~\ref{Fig:energy-NDEigensystem-vs-Bext}(a) do not link different bands and are not topological.  Furthermore, all the Chern numbers for the bands are zero for $B_{\mathrm{ext}} < B_{c,1}$. The lowest energy edge states in Fig.~\ref{Fig:energy-NDEigensystem-vs-Bext}(b) connect bands 5 and 6 but there is no gap between the 5th and 6th bands. The projection of the edge states onto the bulk states of the same energy differs from zero allowing transitions to the system's interior, which would result in a damage of the edge current. Hence these edge states might not be TESs.  In contrast, the upper set of edge states in Fig.~\ref{Fig:energy-NDEigensystem-vs-Bext}(b) that connect bands 6 and 7 are topological  (the Chern number $C_6 = -1$ and $C_7 = 1$ and the other Chern numbers are zero).  The edge states that connect the 6th and 7th bands have regions in $k_x$ that lie within the bandgap.  The bulk-boundary correspondence leads to the emergence of one topologically protected chiral edge state within the bandgap on every edge of the TP material.  There are edge states between the 5th and 6th bands, 6th and 7th, and 7th and 8th bands in Fig.~\ref{Fig:energy-NDEigensystem-vs-Bext}(c).  The Chern numbers $C_6 = C_7 = -1$ and $C_8 = 2$. There is one pair of TESs between the 6th and 7th bands, and two pairs between the 7th and 8th bands.  The latter two pairs are situated between the 7th and 8th bands near the $k_x = 0$ point, but are not seen well in Fig.~\ref{Fig:energy-NDEigensystem-vs-Bext}(c), but are seen in Fig.~\ref{Fig:energy-NDEigensystem-vs-Bext}(d).  The former pair connecting the 6th and 7th bands is located in the interval $0.3 q_0 < |k_x| < 0.4 q_0$. In Fig.~\ref{Fig:energy-NDEigensystem-vs-Bext}(d) there are edge states between all the bands. The Chern numbers  are  $C_6 = 1$, $C_7 = -3$ and $C_8 = 2$, indicating that the edge states between the 6th and 7th bands and the 7th and 8th bands are topological.  A crossing of one pair of edge states between the 6th and 7th bands occurs at $k_x = 0$. It seems plausible that another pair, situated in close proximity to $k_x = \pm 0.4 q_0$, represents a continuation of the edge states observed between the 5th and 6th bands.  The edge states between the 7th and 8th bands contain two pairs.

\section{Non-abelian topological phase}   \label{sec:Non-abelian_phase}

In both Fig.~\ref{Fig:energy-NDEigensystem-vs-Bext}(c) and (d), the coefficients $C_6$, $C_7$ and $C_8$ are non-vanishing and $C_n \ne - C_{n+1}$.  Therefore the TP of the system might be non-abelian. Here, we generalize the concept of an eigenvector frame rotation, as previously defined in reference \cite{Wu_19, Bouhon_20, Jiang_22}, using the projection matrices, and show that indeed the appearance of such a sequence of bands implies non-abelian TP.

We follow the line of argument presented in Refs.~\cite{Wu_19, Bouhon_20, Jiang_22}, wherein the non-abelian topological properties are investigated in momentum space, with a focus on eigenvectors and frame rotations. This concept has been demonstrated to be a valuable tool for the description of one- and two-dimensional non-abelian  topological insulator, as evidenced by the findings presented in Ref.~\cite{Jiang_22} where a specific model of a TI is considered. The system exhibits space-time inversion symmetry, which allows the Hamiltonian to be gauge transformed in such a way that it is real in the momentum space and possesses  real eigenstates. If it is assumed that three bands are involved, the eigenvectors can be mapped onto a three-dimensional space. This allows for an examination of the system's properties by considering the rotational characteristics of the band states as they traverse the Brillouin zone.  The TI studied in Ref.~\cite{Jiang_22} exhibits non-abelian band topology, despite the bands being flat. This is due to the non-abelian quaternion group governing the rotational properties of its bands. It is evident that at least three bands must be considered in order to identify the non-abelian properties, given that rotations in two-dimensional space are abelian.

However, our spin-dependent optical lattice Hamiltonian is not $PT$ symmetric, which results in the eigenvectors being complex. It thus follows that the concept of eigenvector frame rotations must be generalized. We therefore introduce a projection matrix, whose complex elements are defined as the inner product between eigenstates of the system, $\mathcal{U}_{n,n'}({\bf k}) = \langle u_{n,{\bf k}_0} | u_{n',{\bf k}} \rangle$, where $u_{n,{\bf k}}({\bf r})$, according to the Bloch's theorem, is periodic function with period equal to the lattice constant and the integration is over a Wigner-Seitz cell. ${\bf k}_0$ is an arbitrarily chosen reference momentum and the subscript $n$ specifies the band. Note that the eigenstates $|u_{n,{\bf k}_0} \rangle$ are mutually orthogonal. In our case the reference momentum ${\bf k}_0$  is taken to be the  $\Gamma$ point. For an external magnetic field $B_{\rm ext}$ such that $B_{\rm ext} > B_{c,2}$, the sixth, seventh, and eighth bands possess non-zero Chern numbers. The appropriate projection matrix $\mathcal{U}_{n,n'}({\bf k})$ is therefore a $3$$\times$$3$ matrix with $n,n' \in \{6,7,8\}$ and with each row normalized to 1. The properties of the projection matrix $\mathcal{U}({\bf k})$ play a pivotal role in determining the character of the TP.

To demonstrate how this works we compare the structure of the projection matrix for the TPs for $B_{c,1} < B_{\rm ext} < B_{c,2}$ and $B_{c,2} < B_{\rm ext} < B_{c,3}$. The primary difference between these two regions is that in the first region the projection matrices have the form of $2$$\times$$2$ and $1$$\times$$1$ blocks, while in the second they form single $3$$\times$$3$ block matrices. To illustrate this difference we use the Frobenius norm defined as follows: $||\mathcal{U}||^2_F = \sum_{n,n'} |\mathcal{U}_{n,n'}|^2$. Note that, the Frobenius norm (or part thereof) is {\it gauge invariant}. 

\begin{figure}[thb] 
\includegraphics[width=0.90\linewidth]{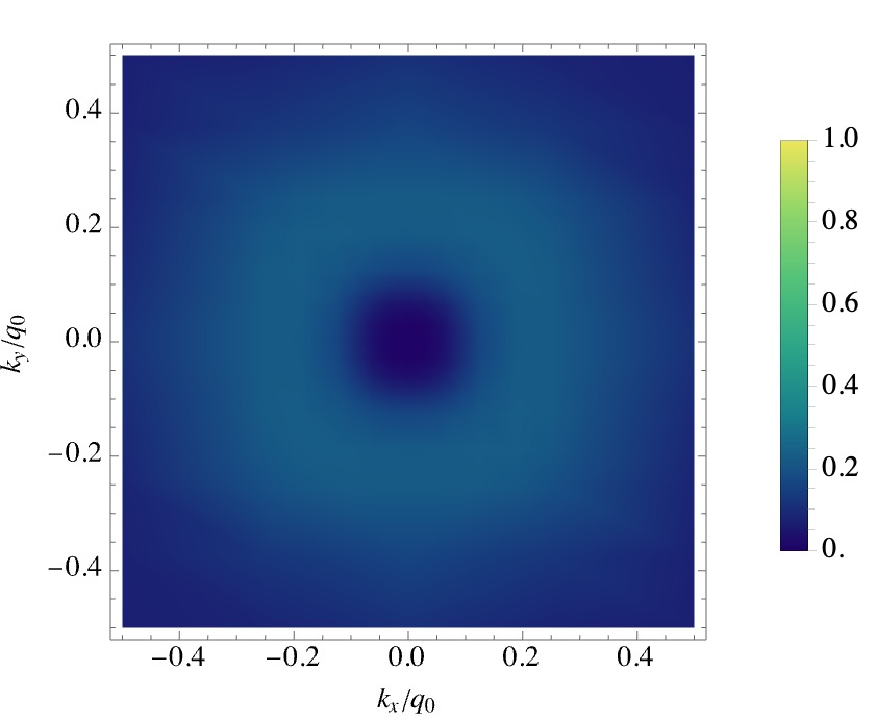}
\includegraphics[width=0.90\linewidth]{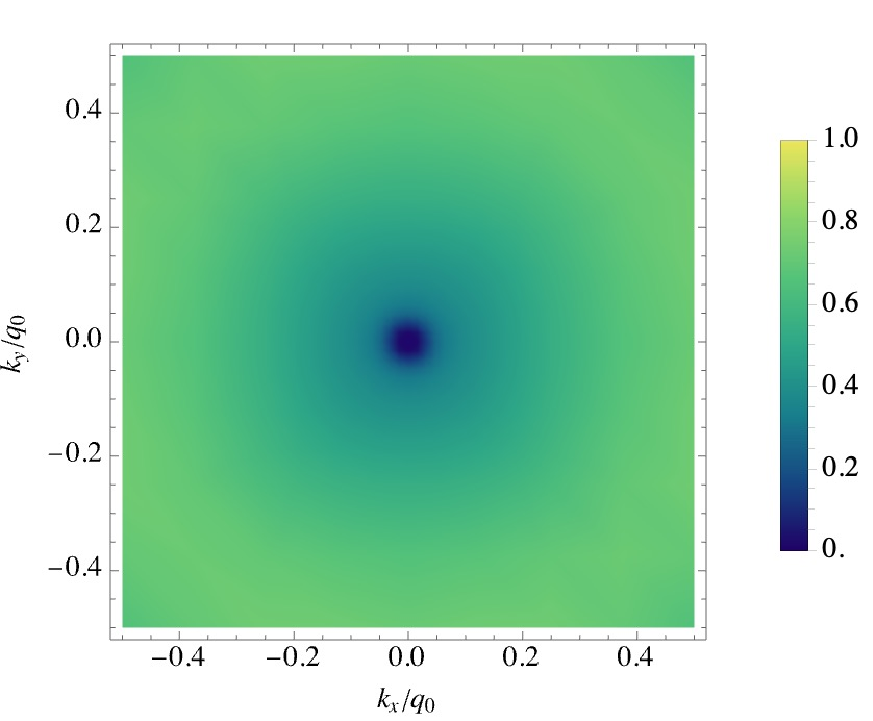}
\caption{Ratio $R$ of the part of the Frobenius norm of the projection matrix to the total Frobenius norm, as indicated in the text. It is plotted as a function of momentum for the following values of the external magnetic field: $B_{c,1} < B_{\rm ext}=0.5 < B_{c,2}$ (top frame) and $B_{c,2} < B_{\rm ext}=0.7 < B_{c,3}$ (bottom frame). For $B_{\rm ext} < B_{c,2}$ value of $R$ is relatively small and the $6$th and $7$th bands separate from the $8$th band,  while for $B_{\rm ext} > B_{c,2}$ $R$ is of order of one and all three bands become entangled and their topology is non-abelian. } 
\label{Fig:off}
\end{figure}

Figure~\ref{Fig:off} depicts the ratio $R$ of part of the Frobenius norm where we include only the following elements: $(n, n') = (6,8)$, $(7,8)$, $(8,6)$, and $(8,7)$ to the total Frobenius norm. In the top frame this norm is small, below $0.2$, which is indicative of the emergence of a two-block structure for $B_{\rm ext}<B_{c,2}$, typical for the topological abelian phase. In contrast, the norm remains large in the bottom frame, indicating that the projection matrix is a full  $3$$\times$$3$ matrix. In the trivial TP, where the external magnetic field $B_{\rm ext} < B_{c,1}$ the projection matrix possess a three $1$$\times$$1$ block structure, which is numerically nearly diagonal.

The observation that in the abelian TP the projection matrix $\mathcal{U}$ has $2$$\times$$2$ and $1$$\times$$1$ block structure implies that $6$th and $7$th bands decouple from the $8$th band. Therefore, the $6$th and $7$th band structure may be considered separately. The eigenstates $|u_{6,{\bf k}_0} \rangle$, $|u_{7,{\bf k}_0} \rangle$, and $|u_{8,{\bf k}_0} \rangle$ are orthonormal and can be represented in a three-dimensional space by mutually orthogonal unit basis vectors $\hat{{\boldsymbol \beta}}$, $\hat{{\boldsymbol \gamma}}$, and $\hat{{\boldsymbol \delta}}$. As the Brillouin zone is traversed, $|\mathcal{U}_{n,n'}({\bf k})| = |\langle u_{n,{\bf k}_0} | u_{n',{\bf k}} \rangle|$, which are gauge independent, can be monitored. The block structure of the projection matrix in the abelian TP (see Fig.~\ref{Fig:off}) indicates that the states $|u_{6,{\bf k}} \rangle$ and $|u_{7,{\bf k}} \rangle$ are primarily confined to the two-dimensional subspace spanned by the vectors $\hat{{\boldsymbol \beta}}$ and $\hat{{\boldsymbol \gamma}}$ Even though the block diagonal structure is not ideal, we can follow a procedure common in quantum information \cite{Reich,Chinol}, where the projection of the unitary evolution in the Hilbert space onto the small subspace is performed and then forced to be again unitary. This can be achieved with the aid of a technique based on the singular value decomposition, which allows for the identification of the closest unitary matrix to a given matrix.

To get insight into the character of state transformations let us consider moving from a point ${\bf k}_1$ in the BZ to the point ${\bf k}_2$. The transformation from the states $|u_{6,{\bf k}_1} \rangle$ and $|u_{7,{\bf k}_1} \rangle$ to $|u_{6,{\bf k}_2} \rangle$ and $|u_{7,{\bf k}_2} \rangle$ is realized using the product operator $\mathcal{U}({\bf k}_2)\, \mathcal{U}^{\dagger}({\bf k}_1)$, which is the $2$$\times$$2$ unitary matrix $\mathcal{\overline{U}}_{n,n'}({\bf k}_1, {\bf k}_2) = \langle u_{n,{\bf k}_1} | u_{n',{\bf k}_2} \rangle$ with $n,n' \in \{6,7\}$. In order to trace the evolution of the band states we consider the gauge-independent quantities $|\mathcal{\overline{U}}_{n,n'}({\bf k}_1, {\bf k}_2)|=|\langle u_{n,{\bf k}_1} | u_{n',{\bf k}_2} \rangle|$. However, the $2$$\times$$2$ unitary matrix after its elements are replaced by their absolute values becomes a real matrix with equal diagonal elements and equal off-diagonal elements. Such matrices form an abelian group, a subgroup of $GL_2(\mathbb{R})$, thereby justifying the designation of the TP.

\begin{figure}[thb] 
\includegraphics[width=0.95\linewidth]{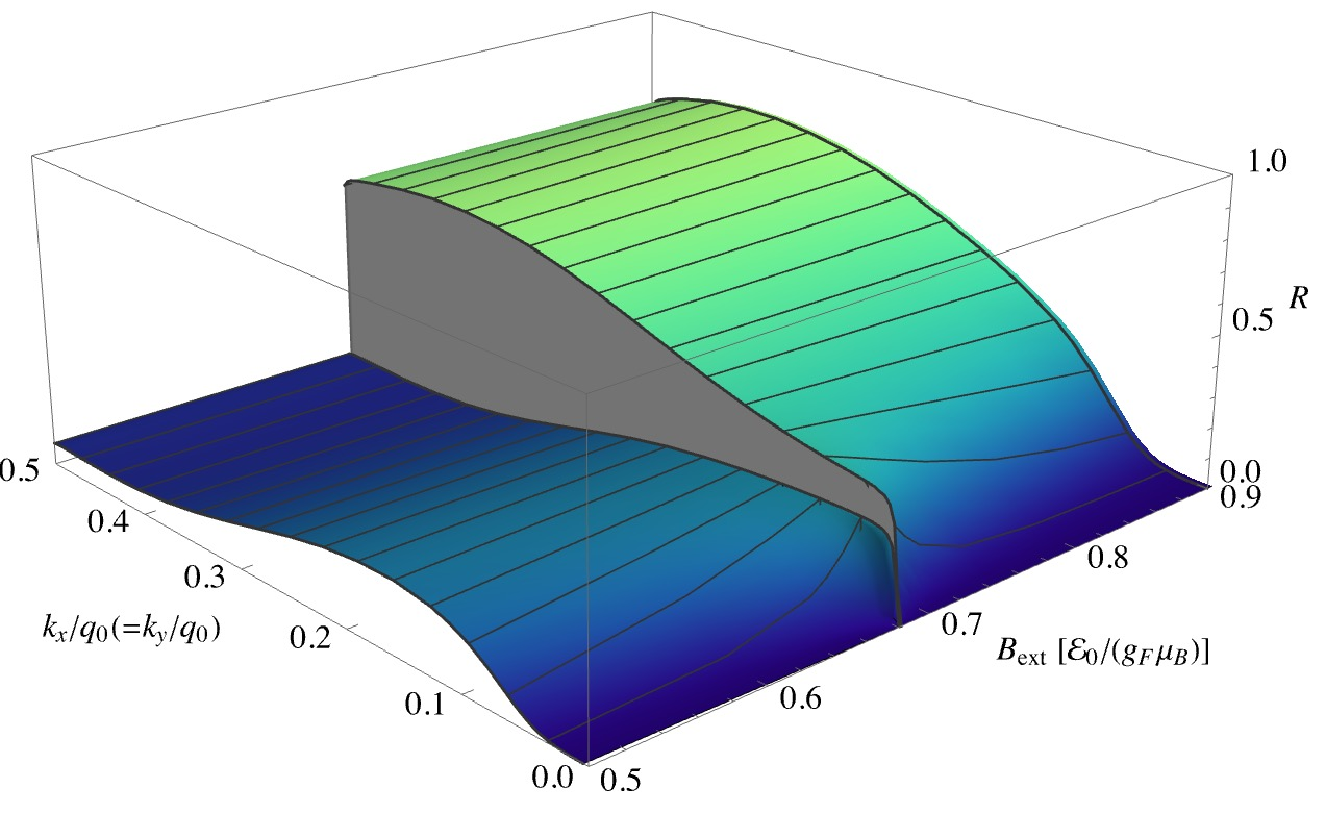}
\caption{Ratio $R$, as in Fig.~\ref{Fig:off}, of the projection matrix, as a function of momentum along the diagonal in the BZ and an external magnetic field $B_{c,1} < B_{\rm ext} < B_{c,3}$. Appearance (or disappearance) of a block diagonal structure has a character of a phase transition. It occurs at $B_{\rm ext} = B_{c,2}$ in correspondence with the TP transition observed at $B_{c,2}$.} 
\label{Fig:ATInonATI}
\end{figure}

In contrast, for an external magnetic field strength exceeding the critical field strength for the second phase transition, $B_{\rm ext} > B_{c,2}$ (Fig.~\ref{Fig:off}, bottom frame), all three bands remain coupled and the band states undergo a full three-dimensional transformation as the system traverses the Brillouin zone. Consequently, for $B_{c,2} < B_{\rm ext} < B_{c,3}$ the TP is non-abelian, since 6th, 7th, and 8th bands transform according to the non-abelian $GL_3(\mathbb{R})$ group. The transition from abelian to non-abelian TP is sharp, as demonstrated in Fig.~\ref{Fig:ATInonATI}, and occurs at the value of the external magnetic field equal to $B_{c,2}$ (see Fig.~\ref{TPT2}).

Similarly, for $B_{c,3}<B_{\rm ext}<B_{c,4}$ the projection matrix exhibits a single $3$$\times$$3$ block structure, thereby demonstrating that the phase is also a non-abelian TP. It should be noted that in the $B_{c,4}<B_{\rm ext}<B_{c,5}$ region, the phase reverts to abelian. This is due to the fact that the sum of components of the Frobenius norm corresponding to the off-diagonal elements: $(6,7)$, $(6,8)$, $(7,6)$, and $(8,6)$ become small with magnitude below 0.1.  This results in the projection matrix assuming the form of a $1$$\times$$1$ by $2$$\times$$2$ block. As a result, the states in the seventh and eighth bands become decoupled from the sixth band.  

In the final phase, $B_{\rm ext}>B_{c,5}$, the Chern numbers are non-zero for the bands 5th, 6th, 7th, and 8th (see Appendix~\ref{append:TPTs}). However, the sum of the Chern numbers for any two successive bands is zero. The projection matrix, which is now a $4$$\times$$4$ matrix, has the form of $2$$\times$$2$ and $2$$\times$$2$ blocks (with the part of the Frobenius norm located off the blocks being smaller than $0.1$). This indicates that the phase is again an abelian phase. Even though four successive bands have nonzero Chern numbers, the system remains in the abelian four-band TP. 

Therefore, the Chern numbers themselves cannot be used to distinguish between abelian and non-abelian TPs. The transition from abelian to non-abelian is indicated rather by a discontinuity of the Frobenius norm of the projection matrix as shown in Fig.~\ref{Fig:ATInonATI}. 

\section{Edge state density, spin density, current density, and spin-current density}  \label{sec:density_curr-density_spin-curr-density}

\begin{figure}[thb] 
\includegraphics[width=0.85\linewidth]{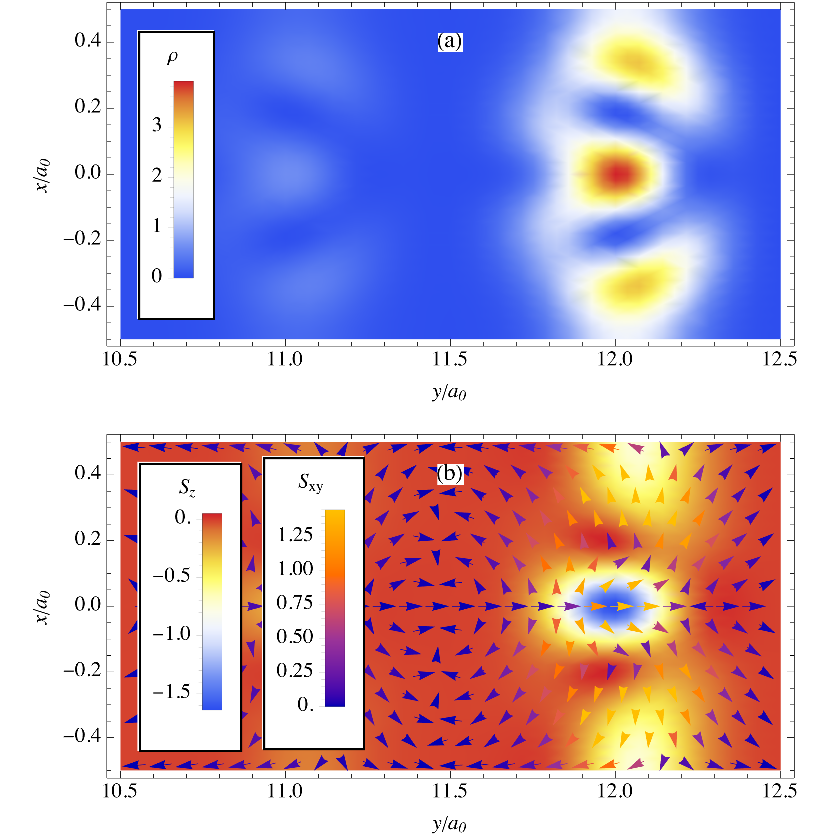}
\caption{ (a) Edge state atom density and (b) spin density for $B_{\mathrm{ext}} = 0.2 \, \mathcal{E}_{0}/(g_F \mu_B)$ and $k_x = -0.2 \, q_0$ and $L_y = 25 a_0$ [see edge state connecting the 6th and 7th bands in Fig.~\ref{Fig:energy-NDEigensystem-vs-Bext}(b)].  The colors of the arrows are shown in the $S_{xy}$ legend, and the background color which gives the $S_z$ component of the spin density is shown in the left legend.}  
\label{atom-density}
\end{figure}

The density of a representative edge state with negative $k_x$ is shown in Fig.~\ref{atom-density}(a).  The density is restricted to the region near the $y = L_y/2$ edge, and the atoms have negative group velocity $v_g(k_x) = \hbar^{-1} d\epsilon/dk_x$.  The nodes in the edge state density are due to the excited state nature of the edge state. Edge states near $y = - L_y/2$ (not shown) have positive $k_x$ and positive $v_g(k_x)$. 

Figure \ref{atom-density}(b) shows the edge state spin density
${\boldsymbol{{\mathcal{S}}}}({\bf r}) = \psi^{\dagger}_{k_x}({\bf r}) {\bf F}  \psi_{k_x}({\bf r})$
(which depends on $k_x$), the arrows show the 2-dimensional vector 
${\boldsymbol{\mathcal{S}}}_{xy}(\mathbf{r}) \equiv (\mathcal{S}_{x}(\mathbf{r}), \mathcal{S}_{y}(\mathbf{r}))$,
and the color of the arrows show the length $\sqrt{ \mathcal{S}_{x}^{2}(\mathbf{r}) +\mathcal{S}_{y}^{2}(\mathbf{r})}$.  
The density plot in Fig.~\ref{atom-density}(b) shows the $z$-component of the spin density vector, $\mathcal{S}_{z}(\mathbf{r})$ which is negative everywhere except where the wave function has nodes (vortices).
The texture of the 2D spin vector ${\boldsymbol{\mathcal{S}}}_{xy}(\mathbf{r})$ is due to the $\mathbf{r}$-dependence of $\mathbf{B}_{\mathrm{fic}}(\mathbf{r})$:
$\mathcal{S}_x({\bf r})$ is an odd function of $x$ since $B_{x, \mathrm{fic}}(\mathbf{r})$
is odd, and $\mathcal{S}_y({\bf r})$ is an even function of $x$, since  $B_{y, \mathrm{fic}}(\mathbf{r})$ is even.  

\begin{figure}[thb] 
\includegraphics[width=0.8 \linewidth]{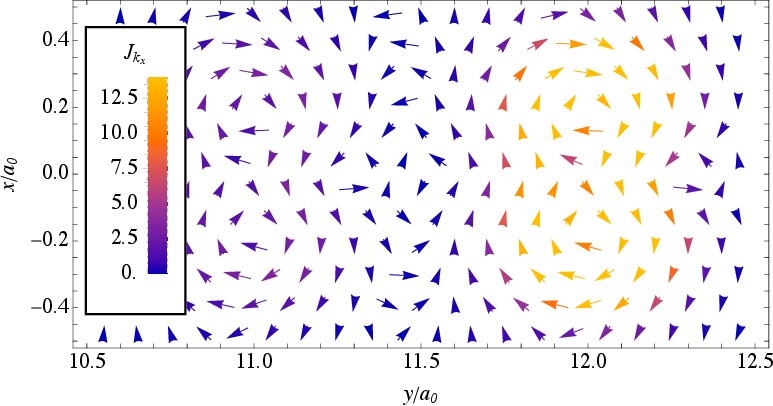}
\caption{Atom current density $\mathbf{J}_{k_x}(\mathbf{r})$ for the parameter values used in Fig.~\ref{atom-density}. Note that the current density flows clockwise, i.e., it has vorticity.}  
\label{Fig:atom-current}
\end{figure}

Figure~\ref{Fig:atom-current} shows the atomic current density $\mathbf{J}_{k_x}(\mathbf{r}) = \tfrac{\hbar}{M} \, \mathrm{Im} \, [ \psi_{k_x}^{\dag}(\mathbf{r}) \, {\boldsymbol \nabla} \psi_{k_x}(\mathbf{r}) ]$, which can be separated into two parts: $\mathbf{J}_{k_x}(\mathbf{r}) = J_{k_x, x}^{\mathrm{av}}(y) \, \hat{\mathbf{x}}+ \mathbf{J}_{k_x}^{\mathrm{vor}}(\mathbf{r})$.  Here $J_{k_x, x}^{\mathrm{av}}(y) = \tfrac{1}{a_0} \int_{-a_0/2}^{a_0/2} J_{k_x, x}(\mathbf{r}) \, dx$ is an `average' current propagating along the edge of the SDOL, and $\mathbf{J}_{k_x}^{\mathrm{vor}}(\mathbf{r})$ which describes a vortex current flow, i.e., a rotational part of the flow.  The net atomic current along the edge can be calculated as $J^{\mathrm{tot}}_{k_x,x} \equiv \int_{-L_y/2}^{L_y/2} J_{k_x, x}^{\mathrm{av}}(y) dy$; the total current is negative and depends on $k_x$ and $J^{\mathrm{tot}}_{k_x,y}$ vanishes for all $k_x$.

\begin{figure}[thb] 
\includegraphics[width=0.8\linewidth]{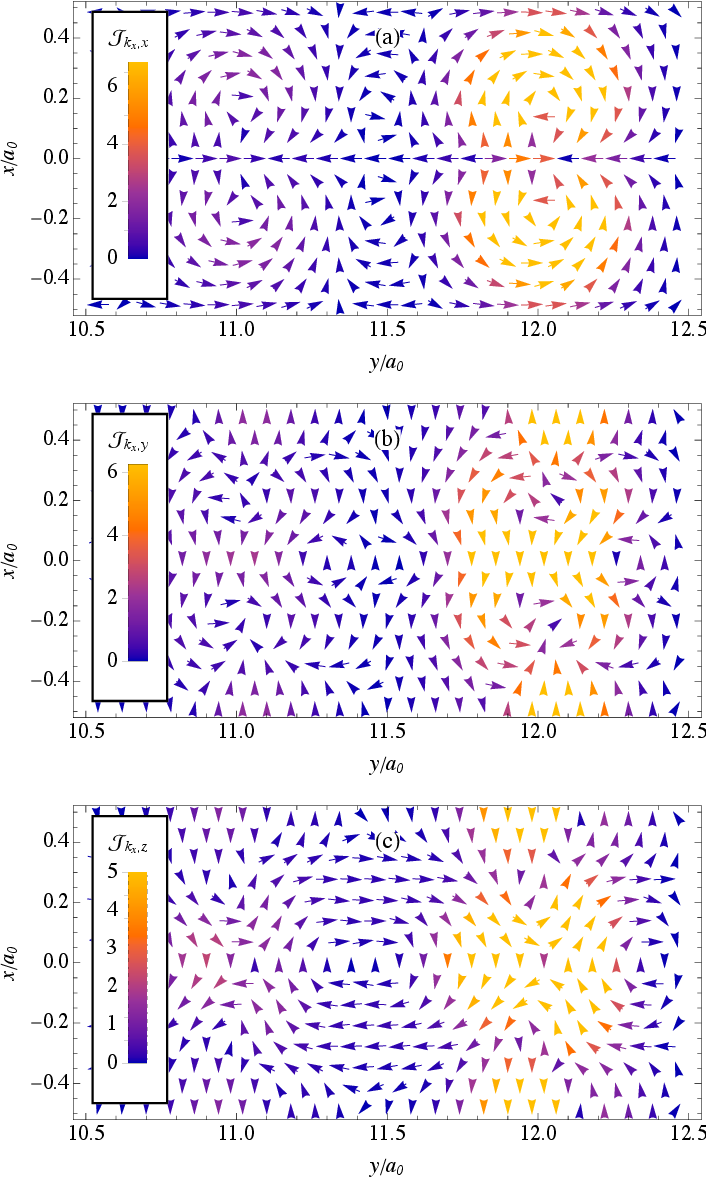}
\caption{Spin-current density for the parameter values used in Fig.~\ref{atom-density}. (a) $\boldsymbol{\mathcal{J}}_{k_x,x}(\mathbf{r})$, (b) $\boldsymbol{\mathcal{J}}_{k_x,y}(\mathbf{r})$, and (c) $\boldsymbol{\mathcal{J}}_{k_x,z}(\mathbf{r})$. The color of the arrows show  $\mathcal{J}_{k_x, \alpha}(\mathbf{r}) = | \boldsymbol{\mathcal{J}}_{k_x, \alpha}(\mathbf{r})|$, where $\alpha = x, y, z$.} 
\label{Fig:spin-current}
\end{figure}

The atomic spin-current density can be denoted by $\mathcal{J}_{k_x,\alpha, \alpha'}(\mathbf{r}) = \tfrac{\hbar}{M} \, \mathrm{Im} \, [ \psi_{k_x}^{\dag}(\mathbf{r}) \, F_{\alpha} \partial_{\alpha'} \psi_{k_x}(\mathbf{r}) ]$.  The subscript $\alpha = x, y, z$, indicates the spin polarization, and the subscript $\alpha' = x, y$, specifies the current propagation direction.  One can decompose the tensor $\mathcal{J}_{k_x,\alpha, \alpha'}(\mathbf{r})$ into three 2D vectors $\boldsymbol{\mathcal{J}}_{k_x,\alpha}(\mathbf{r}) = (\mathcal{J}_{k_x,\alpha, x}(\mathbf{r}), \mathcal{J}_{k_x,\alpha, y}(\mathbf{r}))$ which describe spin-polarized currents and are shown in Fig.~\ref{Fig:spin-current}.

Both $\boldsymbol{\mathcal{J}}_{k_x, y}(\mathbf{r})$ (in Fig.~\ref{Fig:spin-current}(b)) and $\boldsymbol{\mathcal{J}}_{k_x, z}(\mathbf{r})$ (in Fig.~\ref{Fig:spin-current}(c)) spin-current densities have `average' currents along the edge of the SDOL, and `vorticity' currents due to the rotational character of the edge states. The non-vanishing total spin-currents are given by 
$\mathcal{J}_{k_x, y, x}^{\mathrm{tot}} = \int_{-L_y/2}^{L_y/2} \mathcal{J}_{k_x, y, x}^{\mathrm{av}}(y) dy =
-0.526 \, \hbar / (M a_{0}^{2})$ and
$\mathcal{J}^{\mathrm{tot}}_{k_x,z, x} \equiv \int_{-L_y/2}^{L_y/2} \mathcal{J}_{k_x,z, x}^{\mathrm{av}}(y) dy = -0.236 \, \hbar / (M a_{0}^{2})$.

$\boldsymbol{\mathcal{J}}_{k_x, x}(\mathbf{r})$ in Fig.~\ref{Fig:spin-current}(a) exhibits rotational motion and the net spin-current in the $x$-direction vanishes.  It is unexpected that a non-vanishing spin-current exists in the $y$-direction since the atomic current vanishes in the $y$-direction.  To explain why this happens, let us note that the $y$-component of the current in Fig.~\ref{Fig:atom-current} is an odd function of $x$, and thus the net atomic current vanishes in the $y$-direction.  But the $y$-component of the spin-current density in Fig.~\ref{Fig:spin-current}(a) is an even function of $x$, and the net spin-current density in the $y$-direction does not vanish because both the $y$-component of the atomic current and the  $x$-component of the spin density in Fig.~\ref{atom-density} are odd functions of $x$, and their product is even.

\section{Summary and Conclusion}  \label{sec:summary}

We studied the topological properties of cold fermionic Li atoms in a 2D SDOLP in the presence of an external magnetic field perpendicular to the lattice. Both scalar and vector (fictitious magnetic field) potentials are built into the SDOLP, and both magnetic fields break time-reversal symmetry (see Appendix~\ref{append:SDOLP-symmetries}). Topological phases, protected by spatial symmetries, appear upon increasing the external magnetic field.  The calculated Chern numbers for the lowest energy bands ($n \leq 8$) are all zero for $B_{\mathrm{ext}} < B_{c,1}$.   For $B_{\mathrm{ext}} > B_{c,1}$, the Chern numbers take a series of nonzero values as $B_{\mathrm{ext}}$ increases, revealing both abelian and non-abelian topological states and topological phase transitions. For a finite width lattice, we observe edge states in the energy gaps between successive bands, some of which are topologically protected. Thus, the atoms in the SDOLP behave as a topological insulator. The atom current density and spin-current density of the TESs have vorticity and average flow along the edge. We believe that using the SDOL technique opens new possibilities for studying a wide range of topological phenomena in ultracold atomic systems.

\begin{acknowledgments}
MB was supported by the NCN Grant No.~2019/32/Z/ST2/00016 through the project MAQS under QuantERA funded by the European Union's Horizon 2020 research and innovation program under grant No.~731473.  Some numerical results were obtained using Center of University of Białystok computers.
\end{acknowledgments}

\appendix

\section{SDOLP Hamiltonian symmetries}   \label{append:SDOLP-symmetries}

Here we discuss some symmetries of the Hamiltonian in Eq.~(\ref{eq:Hamiltonian}). For the sake of completeness we first discuss the case without the external magnetic field, then we introduce the symmetries for the case with $B_{\mathrm{ext}} \neq 0$.

\subsection{Symmetries for $B_{\mathrm{ext}} = 0$}
  \label{subsec:symmetries-T-reversal}

Consider first symmetries of the system in the absence of
the external magnetic field, that is, $B_{\mathrm{ext}} = 0$.
In this case, the system is not a time-reversal invariant, since
\begin{equation}   \label{eq:T-symmetry}
  T H_{\mathrm{Stark}}(x, y) T^{-1} = H_{\mathrm{Stark}}(-x, -y) \ne H_{\mathrm{Stark}}(x, y),
\end{equation}
where $T = e^{i \pi F_y} K$, and $K$ is the operator for complex conjugation.
The system is not a $P$ invariant, since
\begin{equation}   \label{eq:P-symmetry}
  P H_{\mathrm{Stark}}(x, y) P^{-1} = H_{\mathrm{Stark}}(-x, -y) \ne H_{\mathrm{Stark}}(x, y) .
\end{equation}
But $H_{\mathrm{Stark}}(x, y)$ is a $PT$ invariant,
\begin{equation}   \label{eq:P-symmetry1}
  P T H_{\mathrm{Stark}}(x, y) (P T)^{-1} = H_{\mathrm{Stark}}(x, y).
\end{equation}
Moreover, $H_{\mathrm{Stark}}(x, y)$ is invariant under twofold rotation around the $z$ axis,
\begin{equation}   \label{eq:C_2z}
  C_{2,z} H_{\mathrm{Stark}}(x, y) C_{2,z}^{-1} = H_{\mathrm{Stark}}(x', y') ,
\end{equation}
where $C_{2, z} = e^{i \pi F_z}$ and $(x',y') = (-x,-y)$.
The product of the time reversal and twofold rotation gives
\begin{equation}   \label{eq:C_2z-T}
  (C_{2,z} T) H_{\mathrm{Stark}}(x, y) (C_{2,z} T)^{-1} = H_{\mathrm{Stark}}(x, y) ,
\end{equation}
hence the Hamiltonian is not symmetric under the product operator $C_{2,z} T$.
However, there is fourfold rotational symmetry $C_{4, z} = e^{i \pi F_z/2}$ of the optical lattice,
\begin{equation}   \label{eq:sq:C_4-symmetry-z}
  C_{4, z} H_{\mathrm{Stark}}(x, y) C_{4, z}^{-1} =
  H_{\mathrm{Stark}}(x^\prime, y^\prime) ,
\end{equation}
where $(x^\prime, y^\prime) = (y, -x)$ are the rotated coordinates.  Similarly, there are also two-fold rotational symmetries about the $x$-axis and $y$-axis,
\begin{eqnarray}
  C_{2, x} H_{\mathrm{Stark}}(x, y) C_{2, x}^{-1}
  &=&
  H_{\mathrm{Stark}}(x, -y) ,  \\  \label{eq:sq_C_2-symmetry-x}
  C_{2, y}  H_{\mathrm{Stark}}(x, y) C_{2, y}^{-1}
  &=&
  H_{\mathrm{Stark}}(-x, y) , \label{eq:sq_C_2-symmetry-y}
\end{eqnarray}
generated by $C_{2, x} = e^{i \pi F_x}$ and $C_{2, y} = e^{i \pi F_y}$.

\subsection{Symmetries for $B_{\mathrm{ext}} \neq 0$}
  \label{subsec:symmetries-Bext}

Now let us consider the system in the presence of a finite external magnetic field applied in the $z$-axis, $\mathbf{B}_{\mathrm{ext}} = B_{\mathrm{ext}} \hat{\mathbf{e}}_{z}$.  The Zeeman interaction Hamiltonian $H_Z$ is not time-reversal invariant, $T H_Z T^{-1} \ne H_Z$, it is $P$ invariant, $P H_Z P^{-1} = H_Z$, therefore it is not $PT$ invariant, $(PT) H_Z (PT)^{-1} \ne H_Z$.  Both the Stark interaction Hamiltonian $H_{\mathrm{Stark}}(x,y)$ and the Zeeman interaction Hamiltonian $H_Z$ are symmetric with respect to the transformations $C_{2,z}$ and $C_{4,z}$ but is not invariant under $C_{2,x}$ and $C_{2,y}$ (since both of them flip $B_{\rm ext}$). Therefore, only two- and fourfold rotation around the $z$-axis remain valid symmetries of the Hamiltonian $H_{\mathrm{Stark,Z}}(\mathbf{r}) = H_{\mathrm{Stark}}(\mathbf{r}) + H_Z$, i.e.
\begin{equation}   \label{eq:C_2z-HZ}
  C_{2, z}
  H_{\mathrm{Stark, Z}}(x,y)
  C_{2,z}^{-1} =
  H_{\mathrm{Stark, Z}}( -x, -y ) = H_{\mathrm{Stark, Z}}(x', y')
\end{equation}
with $(x',y') = (-x,-y)$, and
\begin{equation}   \label{eq:C_4z-HZ}
  C_{4, z}
  H_{\mathrm{Stark, Z}}(x,y)
  C_{4,z}^{-1} =
  H_{\mathrm{Stark, Z}}( y, -x ) = H_{\mathrm{Stark,Z}}(x^\prime, y^\prime) \,
\end{equation}
with $(x^\prime, y^\prime) = (y, -x)$.

\section{Chern numbers and topological phase transitions}   \label{append:TPTs}

\begin{figure}[thb] 
\includegraphics[width=7.0cm]{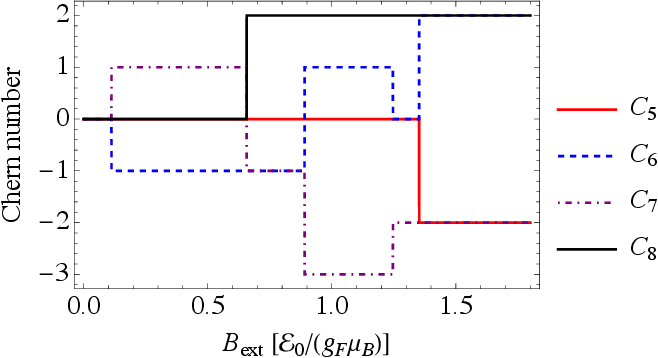}
\caption{The Chern numbers of $5$th, $6$th, $7$th, and $8$th bands as a function of external magnetic field.
The critical points at which the some of the Chern numbers at which the TP transitions occur are
$\{ B_{c,1}, B_{c,2}, B_{c,3}, B_{c,4}, B_{c,5} \} =
\{ 0.11, 0.66, 0.89, 1.25, 1.35 \} \, \mathcal{E}_{0}/(g_F \mu_B)$.}  
\label{CNvsBext}
\end{figure}

In the main text we classify the TPs of a two-dimensional SDOLP. The SDOLP TP in the presence of a finite external magnetic field that is perpendicular to the plane of the  SDOLP breaks time-reversal symmetry. The time-reversal symmetry is broken by both fictitious magnetic field and the external magnetic field. Therefore different TPs can be observed with increasing the strength of the external magnetic field $B_{\mathrm{ext}}$. Figure~\ref{CNvsBext} shows the Chern numbers of $5$th, $6$th, $7$th, and $8$th bands verus external magnetic field strength for $V_0=5\, \mathcal{E}_{0}$ and $B_0=10 \, \mathcal{E}_{0}/(g_F \mu_B)$. $B_{\mathrm{ext}}=1.0\, \mathcal{E}_{0}/(g_F \mu_B)$ corresponds to the external magnetic field equal to 78.8 mG. For lower bands all the Chern numbers equal zero.

\begin{figure}[thb] 
\includegraphics[width=7.0cm]{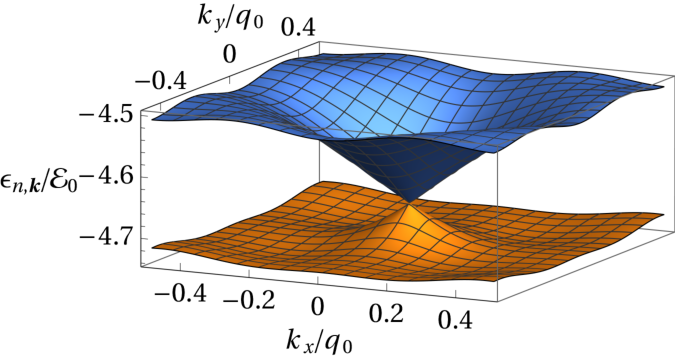}
\caption{The $6$th and $7$th Bloch bands at the external magnetic field $B_{\mathrm{ext}} =  B_{c,1}$. The bands form a Dirac cone at ${\bf k} = {\bf 0}$ so the gap between bands is closed at this value of $B_{\mathrm{ext}}$.}  
\label{DiracCone}
\end{figure}

\begin{figure}[thb] 
\includegraphics[width=7.0cm]{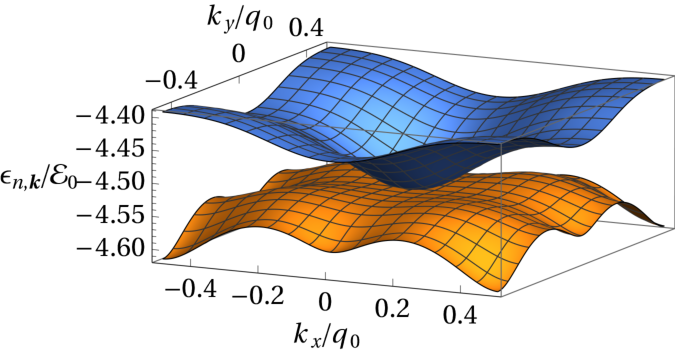}
\caption{The $7$th and $8$th Bloch bands are closed at the external magnetic field
$B_{\mathrm{ext}} = B_{c,2}$ at ${\bf k} = {\bf 0}$ but not in the form of a Dirac cone.}  
\label{TPT2}
\end{figure}

The first TP transition occurs at the external magnetic field $B_{\mathrm{ext}} = B_{c,1}$. The $6$th and $7$th Bloch bands at this external magnetic field touch each other at momentum ${\bf k} = {\bf 0}$. Near the touching point the dispersion of the states is linear, hence the dispersion forms a Dirac cone, see Fig.~\ref{DiracCone}. For $B_{\mathrm{ext}} > B_{c,1}$ the Chern numbers of $6$th and $7$th Bloch bands change from $C_6=C_7=0$ to $C_6=-1$ and $C_7=1$. The Chern numbers of all other bands remain zero. For $B_{c,1} < B_{\mathrm{ext}} < B_{c,2}$ the atoms in a spin-dependent optical lattice potential remain in the phase of an abelian TP. At $B_{\mathrm{ext}} = B_{c,2}$ the second TP transition occurs (see Fig.~\ref{CNvsBext}). Again, the gap is closed at the $\Gamma$ point in the Brillouin zone (Fig.~\ref{TPT2}). However, this time $7$th and $8$th Bloch bands touch each other and the dispersion does not form a Dirac cone. After the second phase transition the Chern numbers of involved bands become $C_7=-1$ and $C_8=2$ an the system of atoms in a spin-dependent optical lattice potential becomes essentially (since $C_6=-1$) a multi-band TP. The system enters the non-abelian TP \cite{Jiang_22}.

\begin{figure}[thb] 
\includegraphics[width=7.0cm]{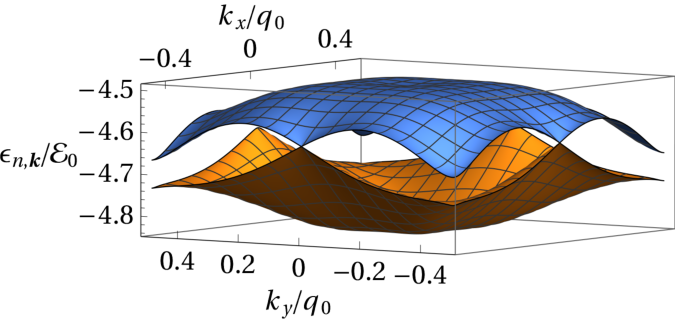}
\caption{The $6$th and $7$th Bloch bands are closed at the external magnetic field
$B_{\mathrm{ext}} = B_{c,3}$ at the centers of the edges of the Brillouin zone.}  
\label{TPT3}
\end{figure}

\begin{figure}[thb] 
\includegraphics[width=7.0cm]{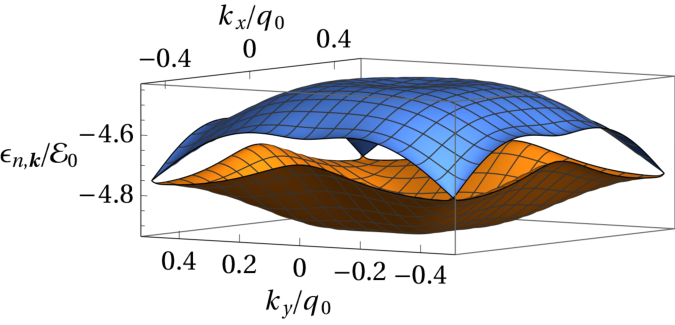}
\caption{The $6$th and $7$th Bloch bands are closed at the external magnetic field
 $B_{\mathrm{ext}} = B_{c,4}$ at the corners of the edges of the Brillouin zone. }  
\label{TPT4}
\end{figure}

\begin{figure}[thb] 
\includegraphics[width=7.0cm]{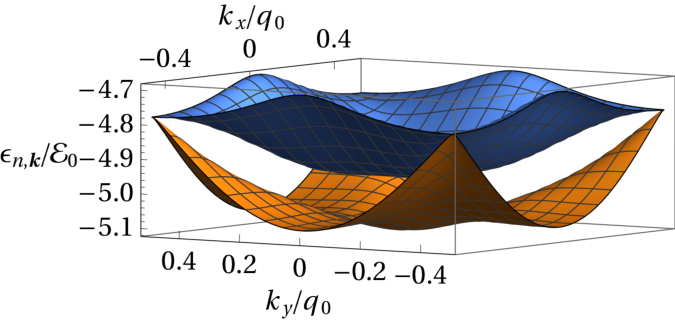}
\caption{The $5$th and $6$th Bloch bands are closed at the external magnetic field $B_{\mathrm{ext}} = B_{c,5}$ at the corners of the Brillouin zone.}  
\label{TPT5}
\end{figure}

Two successive TP transitions involve the $6$th and $7$th bands and the inter-band gap is closed at four points in the BZ. At $B_{\mathrm{ext}} = B_{c,3}$ the bands touch at the centers of the edges of the BZ and the Chern numbers change to $C_6=1$ and $C_7=-3$ (Figs. \ref{TPT3} and \ref{CNvsBext}). The system still remains in the non-abelian TP. However, after the fourth TP transition at $B_{\mathrm{ext}}= B_{c,4}$ (here the gap is closed at the corners of the BZ, see Fig. \ref{TPT4}) the atoms in a spin-dependent optical lattice potential enter back the abelian TP, the only non-zero Chern numbers are those for bands $7$th and $8$th: $C_7=-2$ and $C_8=2$. The last TP transition we studied occurs at $B_{\mathrm{ext}} = B_{c,5}$ and the band gap between $5$th and $6$th bands is closed at the corners of the BZ (see Fig. \ref{TPT5}). After the transition the Chern numbers become $C_5=-2$ and $C_6=2$,  and $C_7=-2$ and $C_8=2$ remain unchanged. Even though four successive bands have nonzero Chern numbers, the system remains in the abelian four-band TP (see Sec.~\ref{sec:Non-abelian_phase}).

\begin{figure}[thb] 
\includegraphics[width=7.0cm]{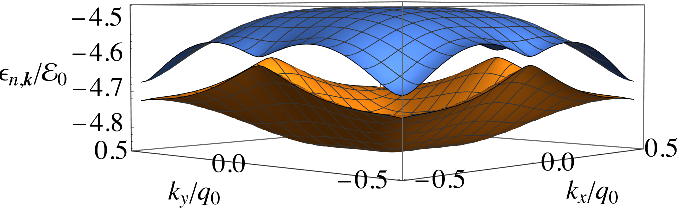}
\caption{The $6$th and $7$th Bloch bands for external magnetic field $B_{\mathrm{ext}} = \mathcal{E}_0/(\mu_B g_F)$.}  
\label{Fig:6_7_gap}
\end{figure}

Figure~\ref{Fig:6_7_gap} shows the energies $\epsilon_{6, \mathbf{k}}$ and $\epsilon_{7, \mathbf{k}}$ of the 6th and 7th Bloch bands respectively.  For all wave vectors $\mathbf{k}$, $\epsilon_{7, \mathbf{k}} > \epsilon_{6, \mathbf{k}}$, i.e., the 6th and 7th bands are separated by a gap.   However, the maximum energy of the 6th band,  $\epsilon_{6, (0, q_0/2)} = -4.615 \, \mathcal{E}_{0}$, is above the minimum energy of the 7th band, $\epsilon_{7, (q_0/2, q_0/2)} = -4.673 \, \mathcal{E}_{0}$.  The finite width of the SDOLP in the strip geometry considered in the main text lifts the translational symmetry in the $y$ direction, hence $k_y$ is not a good quantum number.  Projecting the 3-dimensional image in Fig.~\ref{Fig:6_7_gap} onto the $k_x$-$\epsilon$ plane, we get an overlap of the 6th and 7th bands, hence the band-gap closes, see the red and gold bands in Fig.~\ref{Fig:energy-NDEigensystem-vs-Bext}(d).

\section{Calculating Chern numbers}   \label{append:CN}
To calculate the Chern numbers of the bands for fermionic $^6$Li atoms in the SDOLP we switch from  position space to momentum space. Since $^6$Li atoms are in the hyperfine state with $F=1/2$, the Sch\"odinger equation $H \psi_{n,{\bf k}}({\bf r}) = \epsilon_{n,{\bf k}} \psi_{n,{\bf k}}({\bf r})$, with periodic Hamiltonian having the lattice period $a_0 = 2\pi /q_0$ (see the main text), can be written in a $2\times 2$ matrix form as
\begin{eqnarray}
\left(\begin{array}{cc}
H_+ & W   \\
W^* & H_{-}
\end{array}    \right)
\left(\!\!\begin{array}{c}
\psi_{\frac{1}{2},n,{\bf k}}({\bf r}) \\ \psi_{-\frac{1}{2},n,{\bf k}}({\bf r})
\end{array}  \!\!  \right)
=\, \epsilon_{{\bf k}}
\left( \!\! \begin{array}{c}
\psi_{\frac{1}{2},n,{\bf k}}({\bf r}) \\ \psi_{-\frac{1}{2},n,{\bf k}}({\bf r})
\end{array}  \!\!\ \right)   ,  
\label{HamF12}
\end{eqnarray}
where
\begin{eqnarray}
&& H_{\pm} = -\frac{\hbar^2}{2 M} \nabla^2 + V({\bf r}) \mp \frac{g_F \mu_B}{2} B_{\mathrm{ext}} \, ,  \nonumber \\
&& W = - \frac{g_F\, \mu_B}{2} \left(B_{\mathrm{fic},x}({\bf r}) - i B_{\mathrm{fic},y}({\bf r}) \right) ,
\label{VW12}
\end{eqnarray}
and the wave function $\psi_{n,{\bf k}}({\bf r})$ is a two-component vector $\psi_{n,{\bf k}}({\bf r})=(\psi_{\frac{1}{2},n,{\bf k}}({\bf r}),\psi_{-\frac{1}{2},n,{\bf k}}({\bf r}))^T$.

The wave functions $\psi_{n,{\bf k}}({\bf r})$, which are the solutions of the Sch\"odinger equation in a periodic potential, take the form of a plane wave modulated by a periodic functions, according to the Bloch's theorem, 
\begin{equation} \label{eq:psi}
\psi_{n,{\bf k}}({\bf r}) = e^{i {\bf k} \cdot {\bf r}}\, u_{n,{\bf k}}({\bf r}), 
\end{equation}
where $n$ is a band index and ${\bf k}$ is the momentum vector. Since $u_{n,{\bf k}}({\bf r})$ is periodic with period equal to the lattice constant, it can be expanded in a Fourier series,
\begin{eqnarray}
u_{n,{\bf k}}({\bf r}) = \sum_{{\bf q}} e^{i {\bf q} \cdot {\bf r}} \, \tilde u_{n,{\bf k}}({\bf q})  ,
\label{uFs}
\end{eqnarray}
where ${\bf q}=m_x {\bf q}_1 + m_y {\bf q}_2$ ($m_x$ and $m_y$ are integer numbers) are the reciprocal lattice vectors and $\tilde u_{n,{\bf k}}({\bf q})$ are Fourier coefficients. Similarly, the scalar potential $V({\bf r})$ and the fictitious magnetic field ${\bf B}_\mathrm{fic}({\bf r})$ behave periodically with the lattice constant (see the main text) and can be written as
\begin{eqnarray}
V({\bf r}) &=& \sum_{{\bf q}} V_{{\bf q}} \,e^{i {\bf q} \cdot {\bf r}}   \nonumber \\
W({\bf r}) &=& \sum_{{\bf q}} W_{{\bf q}} \,e^{i {\bf q} \cdot {\bf r}}  \,.
\label{VWFs}
\end{eqnarray}

Inserting the expansions Eqs.~(\ref{uFs}) and (\ref{VWFs}) into Eq.~(\ref{HamF12}) one obtains the Schr\"odinger equation in momentum space, which in what follows will be split into two coupled equations with Fourier coefficients $\tilde u_{\frac{1}{2},n,{\bf k}}$ and $\tilde u_{-\frac{1}{2},n,{\bf k}}$ for the hyperfine spin projections $\frac{1}{2}$ and $-\frac{1}{2}$, respectively:
\begin{eqnarray}
&&\left\{ \frac{\hbar^2 |{\bf k} + {\bf q}|^2}{2 m} - \frac{g_F \mu_B B_{\mathrm{ext}}}{2} \right\}\, \tilde u_{\frac{1}{2},n,{\bf k}}({\bf q})  + \sum_{{\bf q}'}  V_{{\bf q}-{\bf q}'}\, \tilde u_{\frac{1}{2},n,{\bf k}}({\bf q}') 
 \nonumber \\
&& + \sum_{{\bf q}'}  W_{{\bf q}-{\bf q}'}\, \tilde u_{-\frac{1}{2},n,{\bf k}}({\bf q}') = 
\epsilon_{\bf k}\, \tilde u_{\frac{1}{2},n,{\bf k}}({\bf q}) ,
\label{Schinmomsp1}
\end{eqnarray}
and
\begin{eqnarray}
&&\left\{ \frac{\hbar^2 |{\bf k} + {\bf q}|^2}{2 m} + \frac{g_F \mu_B B_{\mathrm{ext}}}{2} \right\}\, \tilde u_{-\frac{1}{2},n,{\bf k}}({\bf q})    \nonumber \\
&& + \sum_{{\bf q}'}  V_{{\bf q}-{\bf q}'}\, \tilde u_{-\frac{1}{2},n,{\bf k}}({\bf q}')  \nonumber \\
&& + \sum_{{\bf q}'}  W^*_{{\bf q}'-{\bf q}}\, \tilde u_{\frac{1}{2},n,{\bf k}}({\bf q}') = 
\epsilon_{\bf k}\, \tilde u_{-\frac{1}{2},n,{\bf k}}({\bf q}) .
\label{Schinmomsp2}
\end{eqnarray}
Here the Fourier coefficients of scalar and vector potentials are given by integrals over an elementary cell
\begin{eqnarray}
&& V_{{\bf q}-{\bf q}'} = 
\frac{1}{a_0^2} \int  e^{-i ({\bf q}-{\bf q}') {\bf r}}\, V({\bf r})\, dx\, dy  \nonumber \\
&& W_{{\bf q}-{\bf q}'} = 
\frac{1}{a_0^2} \int  e^{-i ({\bf q}-{\bf q}') {\bf r}}\, W({\bf r})\, dx\, dy \,.
\label{VUcoef}
\end{eqnarray}
The solutions of Eqs.~(\ref{Schinmomsp1}) and (\ref{Schinmomsp2}), i.e., the eigenstates of the Bloch Hamiltonian $H({\bf k})$, are the Bloch states $|n, {\bf k} \rangle$ in momentum space,
\begin{eqnarray}
|n, {\bf k} \rangle =
\left( \begin{array}{c}
\vdots   \\
\tilde u_{\frac{1}{2},n,{\bf k}}({\bf q})  \\
\tilde u_{-\frac{1}{2},{\bf k}}({\bf q})  \\
\vdots   \\
\end{array}  \!\!\! \right)  \,,
\label{Blochstate}
\end{eqnarray}
where ${\bf q}$ is an infinite set of reciprocal lattice vectors. In the position representation one has
\begin{eqnarray}
u_{\alpha, n, {\bf k}}({\bf r}) \equiv \langle \alpha, {\bf r} |n, {\bf k} \rangle = \sum_{{\bf q}} \tilde u_{\alpha,{\bf k}}({\bf q}) \,e^{i {\bf q} \cdot {\bf r}}   \,,
\label{Blochstate1}
\end{eqnarray}
where $\alpha= \pm 1/2$ is the spin projection on the $z$-axis.

The Chern numbers are found from the formula 
\begin{eqnarray}
C_n = \frac{1}{2 \pi i} \sum_{{\bf k} \in {\rm BZ}} \mathcal{F}_{xy}(n,{\bf k})  \,,
\label{Chern}
\end{eqnarray}
where the Berry curvature $\mathcal{F}_{xy}(n,{\bf k})$ is (see Ref.~\cite{Fukui_05})
\begin{eqnarray}
\mathcal{F}_{xy}(n,{\bf k}) = \ln\, \frac{U_{{\bf k},{\bf k}+\hat{{\bf k}}_x}(n)\, U_{{\bf k}+\hat{{\bf k}}_x,{\bf k}+\hat{{\bf k}}_x+\hat{{\bf k}}_y}(n)}  {U_{{\bf k},{\bf k}+\hat{{\bf k}}_y}(n)\, U_{{\bf k}+\hat{{\bf k}}_y,{\bf k}+\hat{{\bf k}}_x+\hat{{\bf k}}_y}(n)}  \,.   \nonumber \\
\label{FSDOL}
\end{eqnarray}
Here the link variables $U_{{\bf k}',{\bf k}''}(n)$ necessary to calculate the curvature are defined as $U_{{\bf k}',{\bf k}''}(n) \equiv \langle n, {\bf k}' |n, {\bf k}'' \rangle$, where the wavevectors ${\bf k}'$ and ${\bf k}''$ belong to the set $\{{\bf k},{\bf k} + \hat{{\bf k}}_x, {\bf k} + \hat{{\bf k}}_y, {\bf k} + \hat{{\bf k}}_x + \hat{{\bf k}}_y\}$, where $\hat{{\bf k}}$ is a unit vector in direction of one of the Cartesian axes, $k_x$, $k_y$.  Note that ${\bf k}'$ and ${\bf k}''$ are both near the vector ${\bf k}$, i.e., $|{\bf k}'' - {\bf k}| \ll q_0$  and $|{\bf k}'' - {\bf k}| \ll q_0$, where $q_0$ is a reciprocal lattice wavenumber. This yields:
\begin{eqnarray}
U_{{\bf k}',{\bf k}''}(n) &=& \sum_{{\bf q}}  \left[
\tilde u_{\frac{1}{2},n,{\bf k}'}({\bf q})\, \tilde u_{\frac{1}{2},n,{\bf k}''}({\bf q})   \right.     \nonumber \\
&&+\left. \tilde u_{-\frac{1}{2},n,{\bf k}'}({\bf q})\, \tilde u_{-\frac{1}{2},n,{\bf k}''}({\bf q}) \right]  .
\label{Ukpkb}
\end{eqnarray}

We calculate the Chern numbers, Eq.~(\ref{Chern}), numerically on the discretized Brillouin zone, and check the convergence of the results by increasing the size of the numerical grid. However, special care must be taken to satisfy the periodicity of the Bloch Hamiltonian $H({\bf k})$ in momentum space. Let's assume that the Brillouin zone is defined by the set of discrete points $(k_x,k_y)$  in the first Brillouin zone with the values of both momenta on a square grid with period $\Delta k$, i.e., $k_x$ and $k_y \in \{0,\Delta k,2 \Delta k,...,2\pi/a_0-\Delta k\}$. Then, calculating the link variables and the curvature at the points near the boundary of the BZ, when $k_x=2\pi/a_0-\Delta k$ or $k_y=2\pi/a_0-\Delta k$, we need solutions for the points $(k_x=2\pi/a_0,k_y)$ or $(k_x,k_y=2\pi/a_0)$. These points are equivalent to the points $(k_x=0,k_y)$ or $(k_x,k_y=0)$ because of periodicity. Unfortunately, in numerical calculations, the periodicity of the Hamiltonian $H({\bf k})$ is broken because of the finite number of terms used in expansion (\ref{uFs}). The remedy is to neglect the points closest to the edge of the Brillouin zone when calculating the sum in Eq.~(\ref{Chern}), while simultaneously increasing the size of the numerical grid. As a result of this procedure, the sum in Eq.~(\ref{Chern}) approaches an integer number which is the Chern number for the band.

\section{Edge states}   \label{sec:edge-states}

\begin{figure}[tbh] 
{\includegraphics[width=0.48\linewidth]{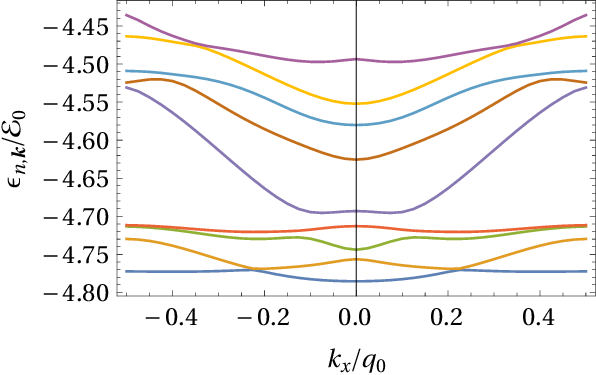}
\includegraphics[width=0.48\linewidth]{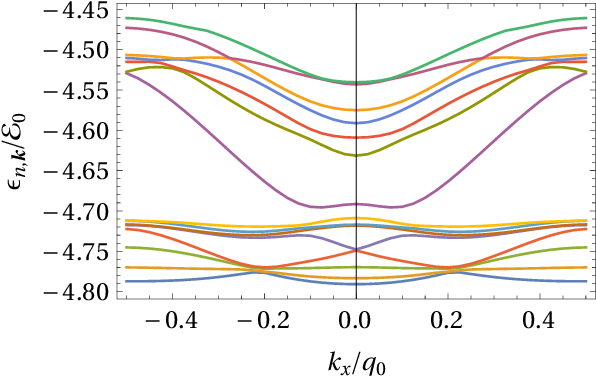}}
{\includegraphics[width=0.48\linewidth]{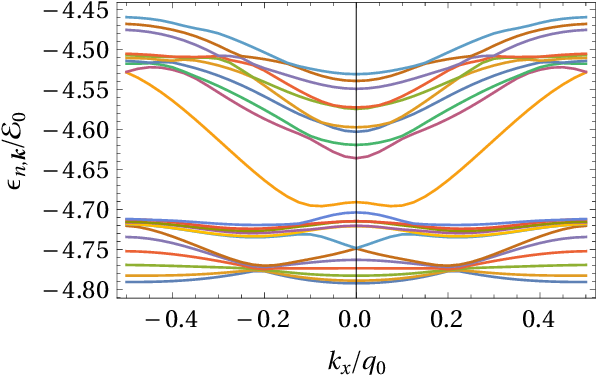}
\includegraphics[width=0.48\linewidth]{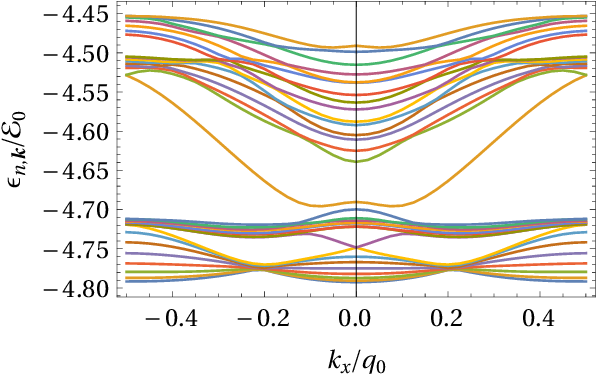}}
\caption{ Energies $\epsilon_{n, k_x}$ of the energy bands 
  $n = 5, 6, 7$, and $8$ 
  for $V_0 = 5 \, \mathcal{E}_{0}$, $B_0 = 10 \, \mathcal{B}_{0}$, and
  for the value of the external magnetic field
  $B_{\mathrm{ext}} = 0.2 \, \mathcal{B}_{0}$.
Successive frames (from left to right and top to bottom) show the spectra for finite width SDOLP strips having $3$, $5$, $7$, and $9$ elementary lattice period.} 
\label{Fig:av-cross}
\end{figure}

Topological edge states are symmetry-protected \cite{Hasan_Kane_10,Qi_Zhang_11}. In the discussion of the SDOLP in the main text (see also Appendix~\ref{append:SDOLP-symmetries}), we point out that there is no time-reversal symmetry, but there are other symmetries, such as two- and fourfold rotational symmetries, $C_{2,z}$ and $C_{4, z}$. For an abelian TP \cite{Hasan_Kane_10}, a pair of edge states propagating on opposite edges has a degeneracy point, i.e., a value of $k_x$ where they have the same energy. For a finite strip width with topological edge states, there is a gap at the degeneracy point, hence the edge states are not topological, but the gap decreases exponentially with increasing width of the strip \cite{Asboth}.  The degeneracy point can be clearly seen (because the width is sufficiently large)  in Fig.~\ref{Fig:energy-NDEigensystem-vs-Bext}(b) between the 5th and 6th bands at $k_x=0$ and between the 6th and 7h bands at the edge of the Brillouin zone. This is illustrated in Fig.~\ref{Fig:av-cross} in the upper left frame, for a SDOLP with a strip-width of $3$ lattice periods, where the lattice period is $a_0 = 2\pi /q_0$.  In this figure one can easily see the gap between the edge states connecting 5th and 6th bands at $k_x=0$, as well as the one occurring at the edge of the Brillouin zone for the states between the 6h and 7th bands. Both avoided crossings disappear when the width of the SDOLP is increased (see Fig.~\ref{Fig:av-cross} for increasing number of cells).

In Figs.~\ref{Fig:energy-NDEigensystem-vs-Bext}(c) and \ref{Fig:energy-NDEigensystem-vs-Bext}(d) there are two pairs of edge states between the 7th and 8th bands, both of which are located near the $k_x=0$ point. One of them has no degeneracy point, and the other has degeneracy point at $k_x=0$. These edge states are topological. For non-abelian topological edge states without a degeneracy point the finite width of the strip doesn't destroy the nature of the TES. For non-abelian topological edge states with a degeneracy point, similar to the abelian case, the gap opens when the width of a strip becomes smaller.

\end{document}